\documentclass[prl,twocolumn,groupedaddress,superscriptaddress,showpacs,floatfix]{revtex4-1}
\usepackage{amsfonts}

\usepackage{graphicx}
\usepackage{dcolumn}
\usepackage{bm}
\usepackage{amsmath}
\usepackage{amssymb}
\usepackage{revsymb4-1}
\usepackage{hhline}
\usepackage[normalem]{ulem}

\usepackage{color}

\usepackage{hyperref}
\hypersetup{colorlinks=true,citecolor={blue},linkcolor={blue},urlcolor={blue}}

\definecolor{ygr}{RGB}{199,0,0}
\newcommand{\mi}{\mathrm{i}}
\newcommand{\oht}{\hat}

\begin{document}

\title{\emph{Zitterbewegung} of exciton-polaritons}

\author{E. S. Sedov}
\email[Electronic address: ]{evgeny\_sedov@mail.ru}
\affiliation{School of Physics and Astronomy, University of Southampton, SO17 1NJ Southampton, United Kingdom}
\affiliation{Department of Physics and Applied Mathematics, Vladimir State University named after A. G. and N. G. Stoletovs, Gorky str. 87, 600000, Vladimir, Russia}
\author{Y. G. Rubo}
\affiliation{Instituto de Energ\'{\i}as Renovables, Universidad Nacional Aut\'onoma de M\'exico, Temixco, Morelos 62580, Mexico}
\affiliation{Center for Theoretical Physics of Complex Systems, Institute for Basic Science (IBS), Daejeon 34051, Republic of Korea}
\author{A. V. Kavokin}
\affiliation{CNR-SPIN, Viale del Politecnico 1, I-00133, Rome, Italy}
\affiliation{School of Physics and Astronomy, University of Southampton, SO17 1NJ Southampton, United Kingdom}
\affiliation{Spin Optics Laboratory, St.\ Petersburg State University, Ul'anovskaya 1, Peterhof, St.\ Petersburg 198504, Russia}

\begin{abstract}
Macroscopic wave packets of spin-polarized exciton-polaritons in two-dimensional microcavities experience the \textit{zitterbewegung}, the effect manifested by the appearance of the oscillatory motion of polaritons in the direction normal to the initial propagation direction.
The oscillating trajectories of exciton-polaritons are adjustable by the control parameters: the splitting of the longitudinal and transverse exciton-polariton modes, the wave vector and the width of the resonant cw pump.
Our theoretical analysis supported by the numerical calculations allowed to optimize values of the control parameters suitable for a direct experimental observation of the \textit{zitterbewegung} effect.
\end{abstract}

\date{\today}

\maketitle

\begin{figure}
\includegraphics[width=0.99\linewidth]{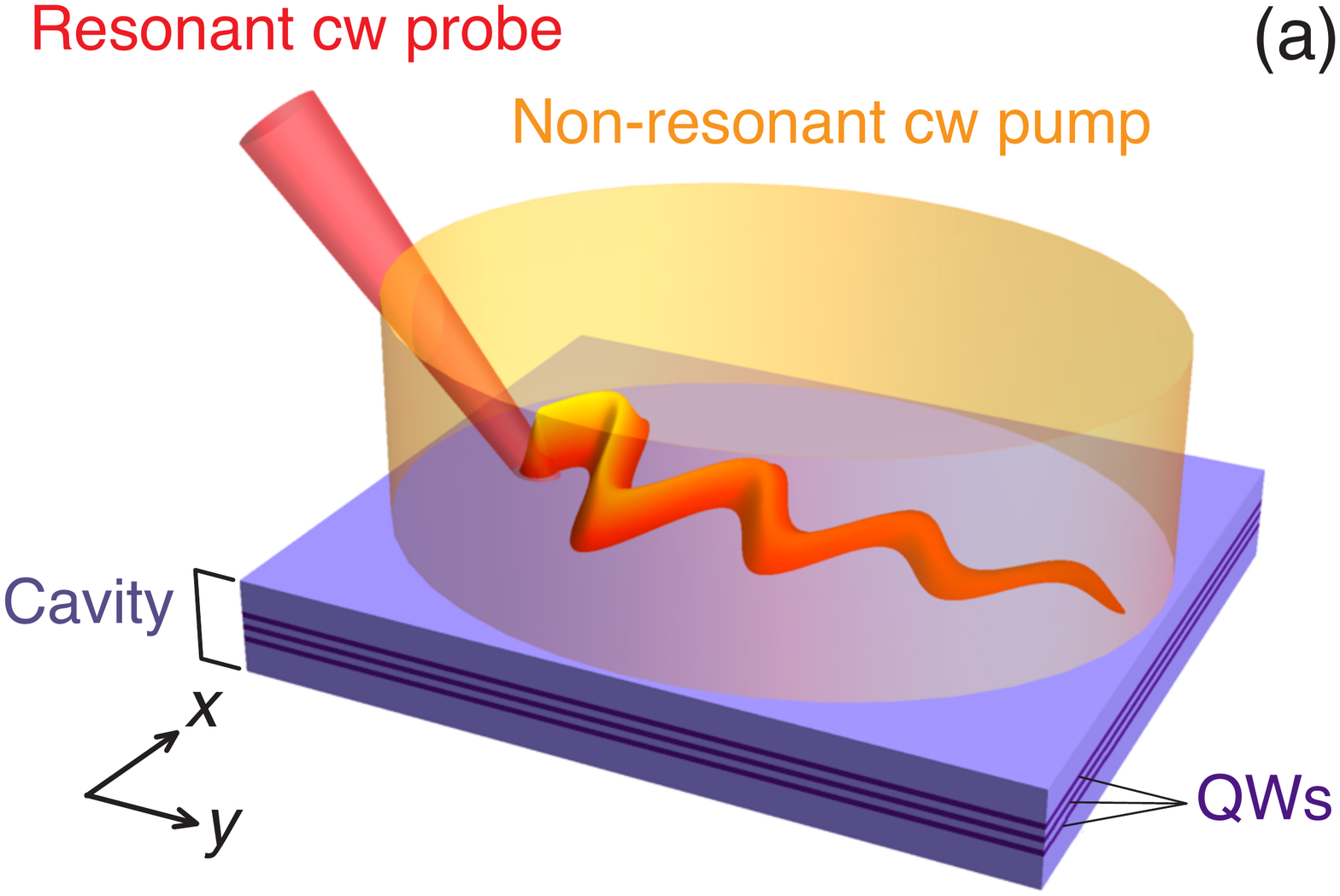}
\includegraphics[width=0.99\linewidth]{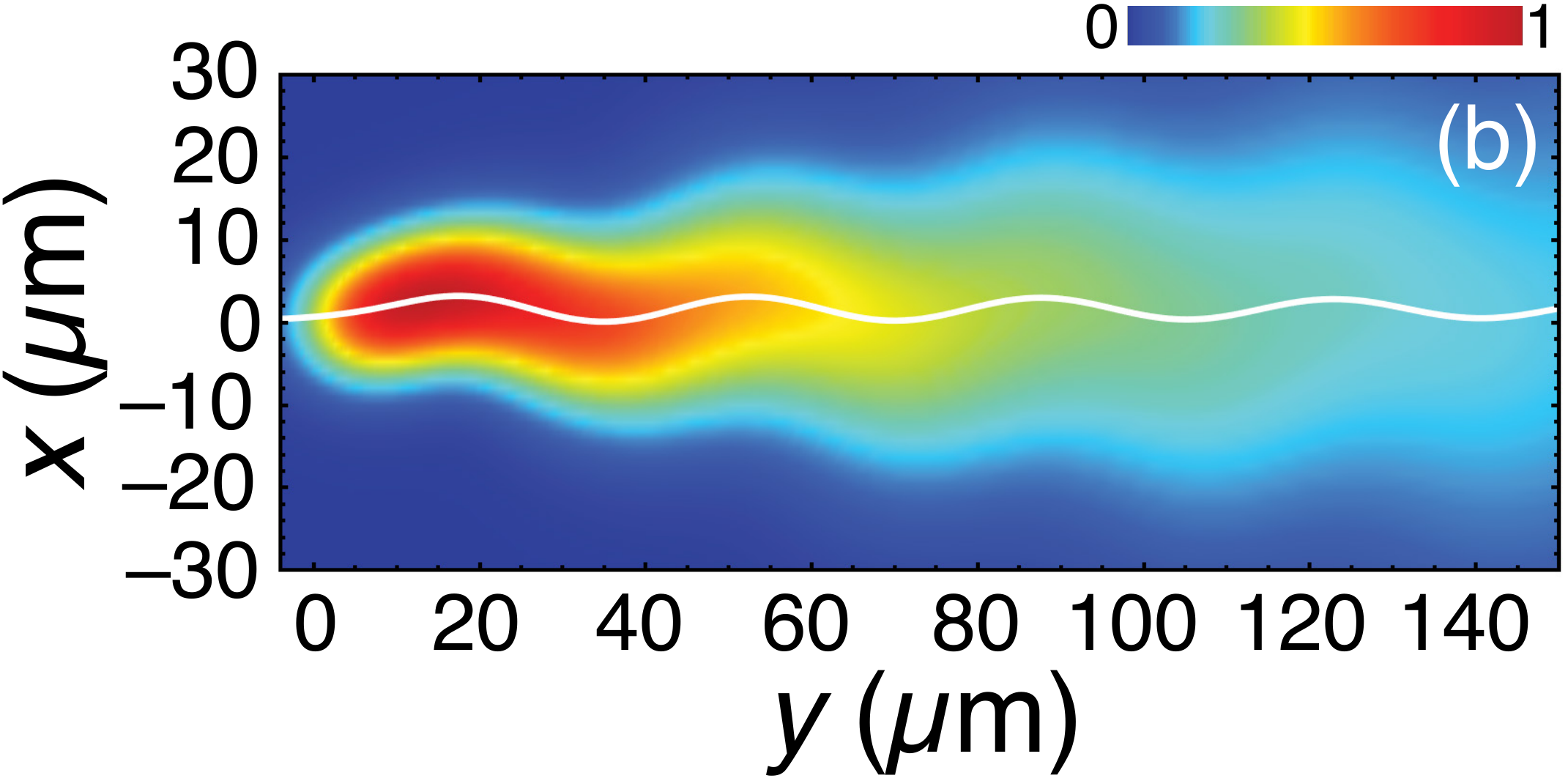}
\caption{ \label{FIGscheme}
(Color online)
(a) Schematic of the considered experimental configuration.
The exciton-polariton condensate is excited by a resonant cw probe in a microcavity with embedded QWs in the presence of a sub-threshold non-resonant cw pump.
(b) The spatial distribution of the polariton condensate intensity in the steady state. 
Values of the parameters used for modelling are given in~\cite{ValuesOfTheParameters}.
}
\end{figure}

The \textit{zitterbewegung} is a contre-intuitive trembling motion of a Dirac particle around the ballistic trajectory~\cite{Nature463682010}.
First dicussed by Schr\"{o}dinger for free electrons~\cite{SitzPressAkadWissPhysMath244181930} and latter predicted for the electrons with the Rashba and Dresselhaus spin-orbital coupling in Refs.~\cite{PhysRevLett.94.206801,PhysRevB.73.085323,PhysRevB.75.205314}, the effect has never been observed experimentally.
The trembling of trajectories in the systems with spectra possessing energy gaps has been observed in Bose-Einstein condensates of ultracold atoms~\cite{PhysRevA.88.021604,0295-5075-83-5-54002}, ions~\cite{Nature463682010}, photonic lattices~\cite{PhysRevLett.105.143902}, graphene and carbone nanotubes~\cite{PhysRevB.76.195439} and even in ordinary sonic crystals (acoustic zitterbewegung)~\cite{WANG20104933}. 
The similar trajectory effect for light beams in specially designed photonic structures have been recently predicted~\cite{NaturePhotonics2748EP22008,PhysRevA.75.053821}.
In this Letter, we report on the \textit{zitterbewegung} of the non-degenerate in pseudospin condensate of exciton polaritons in a 2D microcavity containing an embedded set of quantum wells.
Semiconductor microcavities in the strong exciton-photon coupling regime are host for a number of remarkable bosonic effects such as polariton lasing~\cite{PhysRevLett.98.126405} and superfluid behaviour~\cite{NaturePhysics58052009}.
Cavity polaritons also exhibit strong spin-orbit interaction effects including the optical spin Hall effect~\cite{NaturePhysics36282007}.
The experiment geometry that would allow for the observation of the predicted effect is schematically shown in Fig.~\ref{FIGscheme}(a).
The splitting of the longitudinal and transverse polariton modes (LT splitting) in a microcavity gives rise to the pseudospin precession of polaritons in microcavities~\cite{PhysRevLett.95.136601,ScientificReports797972017}.
In this context, the experimental geometry we consider is similar to one used by Amo \textit{et al.}~\cite{NaturePhysics58052009} to demonstrate the superfluid behaviour of exciton polaritons.
The difference is in the specific choice of the incident probe polarization (circular polarization), incidence angle and, most importantly, a carefully selected LT splitting.

The propagation of polarized exciton-polaritons in the plane of the microcavity is described by the following effective Hamiltonian:
\begin{equation}
  \label{EffHamiltonian}
  \oht{H} = \frac{\hbar^2 \oht{\mathbf{k}}^2}{2 m^*} + \hbar \oht{\boldsymbol{\Omega}} \cdot \oht{\mathbf{S}},
\end{equation}
where $m^{*}=\left.2m_lm_t\right/(m_l+m_t)$, with $m_l$ and $m_t$ being the effective masses of longitudinal (transverse-magnetic) and transverse (transverse-electric) polariton modes, respectively,
$\oht{\mathbf{k}} = (\oht{k}_x, \oht{k}_y) = (-i\partial _x, -i \partial _y)$ is the two-dimensional polariton wave vector.
The polariton quantum state is described by the spinor $|\psi\rangle=(\psi_{+},\psi_{-})^{\mathrm{T}}$, and we use the three-dimensional spin operator
$\oht{\mathbf{S}}=\frac{1}{2}\oht{\boldsymbol{\sigma}}$ acting on $|\psi\rangle$, where $\hat{\sigma}_{x,y,z}$ are the Pauli matrices. 
Spin precession of noninteracting polaritons around the effective field 
$\oht{\boldsymbol{\Omega}}=\big[\Delta_{\text{LT}}(\oht{k}_x^2-\oht{k}_y^2),2\Delta_{\text{LT}}\oht{k}_x\oht{k}_y,0\big]$ 
is due to the longitudinal-transverse splitting of the eigenmodes of the Hamiltonian \eqref{EffHamiltonian}, where 
$\Delta_{\mathrm{LT}}=(\hbar/2)(m_l^{-1}-m_t^{-1})$ is the splitting constant~\cite{PhysRevLett.95.136601}.
The position operators $\oht{x}(t)$ and $\oht{y}(t)$ evolve according to the Heisenberg equations
\begin{subequations}
\label{HeisEqs}
\begin{align}
  \frac{d\oht{x}}{dt}=&\frac{\hbar}{m^{*}}\oht{k}_x+2\Delta_{\mathrm{LT}}(\oht{k}_x\oht{S}_x+\oht{k}_y\oht{S}_y), \\
  \frac{d\oht{y}}{dt}=&\frac{\hbar}{m^{*}}\oht{k}_y-2\Delta_{\mathrm{LT}}(\oht{k}_y\oht{S}_x-\oht{k}_x\oht{S}_y),
\end{align}
\end{subequations}
accompanied by the equation for the precession of the pseudospin operator 
$d\oht{\mathbf{S}}/dt=\oht{\boldsymbol{\Omega}}\times\oht{\mathbf{S}}$.

We consider the initial polariton wave function in the form $|\Psi(\mathbf{r})\rangle=f(\mathbf{r})|\varphi\rangle$, where 
$f(\mathbf{r})=(2\pi)^{-1}{\int}F(\mathbf{k}-\mathbf{k}_0){\exp}(\mi\mathbf{k}\mathbf{r})d\mathbf{k}$ 
describes the spatial distribution and $|\varphi\rangle$ defines the polarization.
In the case of a ballistic polariton condensate moving in $y$ direction, $\mathbf{k}_0||\widehat{\mathbf{y}}$, the effect of \emph{zitterbewegung} consists in the appearance of the non-zero displacement $x(t)=\langle\oht{x}(t)\rangle$ and the non-zero velocity $v_x(t)={\langle}d\oht{x}/dt\rangle$ in the initial direction perpendicular to the direction of propagation.
In the case of the initial circular polarization, $|\varphi\rangle=(1,0)^{\mathrm{T}}$, the displacement is found to be
\begin{equation}
\label{CircDisplacementGeneral}
x(t)=\int\frac{k_y}{k^2}|F(\mathbf{k}- k_{y0})|^2\left[1-\cos(\Delta_{\mathrm{LT}}k^2 t)\right]d\mathbf{k}.
\end{equation}
For a wide in space polariton wave packet with the size $d\gg\lambda=2\pi/k_0$, $F(\mathbf{k}-k_{0})$ is a narrow peak centered at~$\mathbf{k}_0$.
The $x$-displacement as a function of $y$-position of the packet can be found from Eq.~\eqref{CircDisplacementGeneral}:
\begin{equation}
  \label{xONyEQ}
  x(y)=k_{0}^{-1}\left[1-\cos(\Delta_\mathrm{LT}k_0m^*y/\hbar)\right],
\end{equation}
which is similar to the corresponding limit for the electron wave packet in the presence of Rashba spin-orbit coupling \cite{PhysRevLett.94.206801}.
Equation~\eqref{xONyEQ} describes an undamped in space, but very low-amplitude \emph{zitterbewegung}: the maximal displacement of the packet is about the wave length.
As a result, Eq.\ \eqref{CircDisplacementGeneral} predicts the strongest \emph{zitterbewegung} for the polariton packet size comparable to its wavelength, $d\sim\lambda$.

To analyse the \emph{zitterbewegung} in the realistic experimental conditions, and to identify the best structure and experimental set-up characteristics for the observation of the effect, we use the macroscopic mean-field model describing the dynamics of the exciton-polariton condensate.
The model is based on the generalized Gross-Pitaevskii equation for the components of the spinor 
$| \Psi (\mathbf{r}, t) \rangle=(\Psi_{+}(\mathbf{r},t),\Psi_{-}(\mathbf{r},t))^\mathrm{T}$~\cite{PhysRevLett.97.066402,PhysRevLett.109.036404}
\begin{eqnarray}
\label{GPE}
  \mi\hbar\frac{\partial\Psi_{\pm}}{{\partial}t}&=&\left[-\frac{\hbar^2}{2m^{*}}\nabla^2+\alpha_{1}|\Psi_{\pm}|^2+\alpha_{2}|\Psi_{\mp}|^2\right. 
  \nonumber \\
  &+&\alpha_R n_{\pm} + \left. \frac{\mi\hbar}{2} \left( R n_{\pm} - \gamma _C \right) \right] \Psi _{\pm} \nonumber  \\
  &-& \frac{\hbar \Delta_{\text{LT}}}{2} \left( \frac{\partial}{\partial x} \mp \mi \frac{\partial}{\partial y} \right)^2 \Psi _{\mp} + \mi E_{p\pm} (\mathbf{r}) e^{-\mi\omega_\mathrm{p}t} \quad
\end{eqnarray}
coupled to the rate equation describing the evolution of the exciton reservoir density $n_{\pm}(\mathbf{r},t)$:
\begin{equation}
  \label{ReservoirRateEq}
  \frac{\partial n_{\pm}}{\partial t} = P_{\pm} - (\gamma_R + R|\Psi_{\pm}|^2) n_{\pm}.
\end{equation}

In Eqs.~ \eqref{GPE} and \eqref{ReservoirRateEq}, $\alpha_1$ and $\alpha_2$ are the interaction constants for polaritons in the triplet configuration (parallel spins) and in the singlet configuration (opposite spins), respectively,
$\alpha_R$ describes the interaction of the condensate polaritons and the reservoir excitons,
$R$ is the stimulated scattering rate describing the particle exchange between the condensate and the reservoir, and
$\gamma _C$ and $\gamma _R$ are the loss rates of the condensate polaritons and the reservoir excitons, respectively.
In the absence of the in-coming flow of particles from the reservoir, the polariton condensate would only live a short time comparable with the polariton lifetime, which is typically of the order of $\gamma_C^{-1}\sim10\,\mathrm{ps}$.
In this case the condensate would propagate only over tens of micron.
We note, however, at this point that recently substantially longer polariton lifetimes $\gamma_C^{-1}\sim200\,\mathrm{ps}$ and much longer travel distances (hundreds of microns to millimeters) have been achieved~\cite{Appl.Phys.Lett.110.211104,PhysRevLett.118.016602}.
In order to achieve a strong \emph{zitterbewegung} effect at long distances, we propose to excite polaritons resonantly by a cw probe beam 
$E_{p\pm}(\mathbf{r})$ with $\omega_\mathrm{p}$ being the frequency of the resonant pump relative to the minimum of the lower polariton branch.
The wave vector of the probe beam is preserved by the polariton condensate resulting in appearance of the polariton current.
The power of the probe is assumed to be low enough to enable us considering the system in the linear regime when interaction terms in Eq.~\eqref{GPE} are not important and the analytical predictions of Eqs.~\eqref{CircDisplacementGeneral}--\eqref{xONyEQ} should be valid.
We also introduce a non-resonant spatially homogeneous spin-polarized cw pump $P_{\pm}$ in the ``dark regime'', $P_{\pm}<P_{\mathrm{th}}$, where $P_\mathrm{th} = \left. \gamma _C \gamma _R \right/ R$ is the threshold pump~\cite{NaturePhotonics43612010}.
The pump and the probe are considered to be linearly and circularly polarized, respectively.

We solve numerically Eqs.~\eqref{GPE} and~\eqref{ReservoirRateEq} with an initial condition of zero polariton density everywhere in the cavity plane.
The time evolution of the system is calculated  in order to find the steady state regime that is usually established on a time-scale of $300\div400$~ps.
We plot the polariton space distributions of the polariton density $I(\mathbf{r}) = |\Psi _{+} (\mathbf{r})|^2 +|\Psi _{-} (\mathbf{r})|^2 $ in the steady state at $t=700$~ps in Fig.~\ref{FIGscheme}(b).
In simulations, we use the paramenters of Ref.~\cite{ValuesOfTheParameters}.
The \emph{zitterbewegung} results in a trembling spatial oscillatory motion of the polariton condensate center of mass.
The center-of-mass steaty-state trajectory $x(y) $ found as  $x(y) = \left. \int I(\mathbf{r}) x dx \right / \int I (\mathbf{r}) dx$ is shown by the white curve in~Fig.~\ref{FIGscheme}(b).

\begin{figure}
\includegraphics[width=0.99\linewidth]{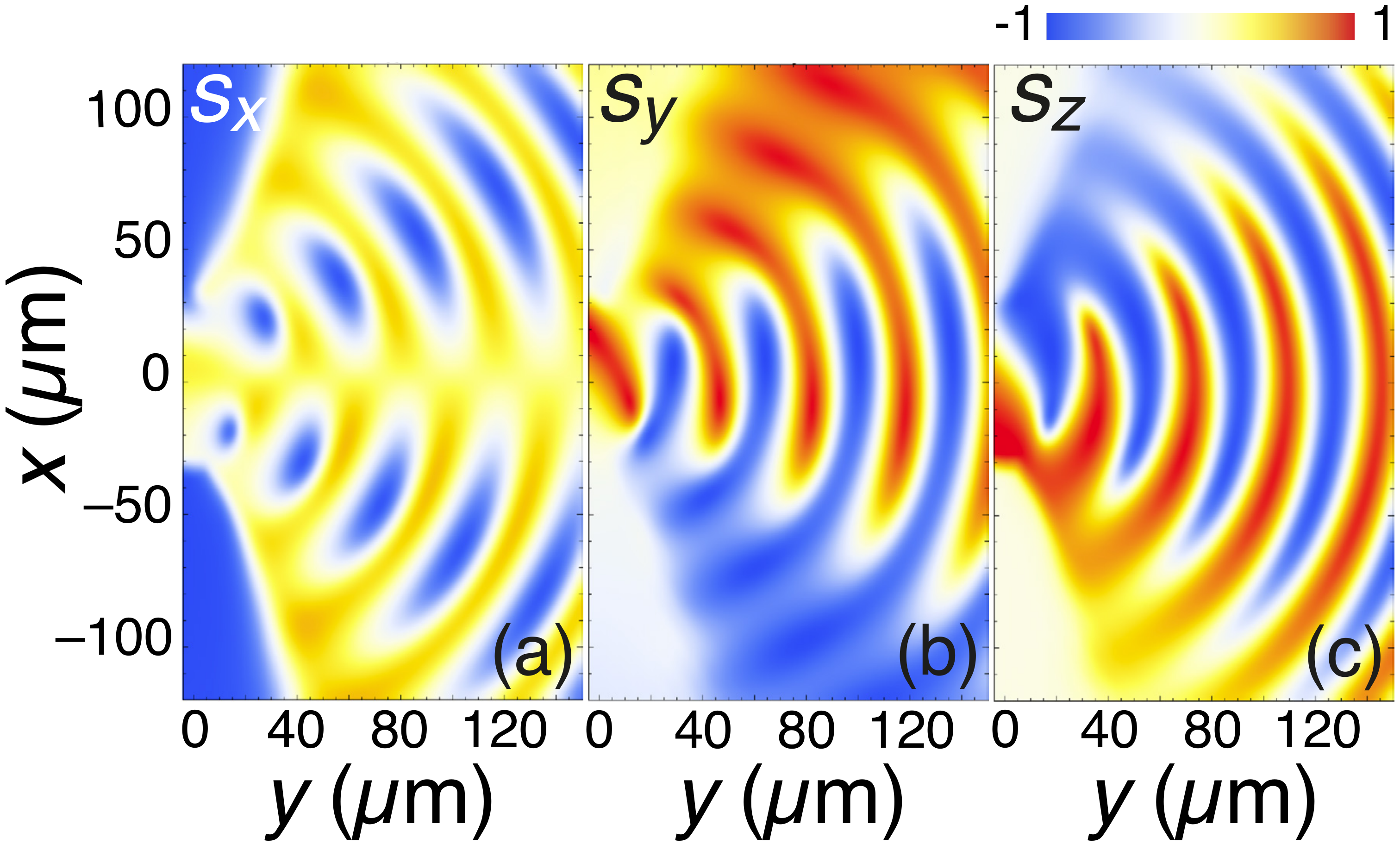}
\caption{ \label{FIGpolarizations}
(Color online)
The effect of the \textit{zitterbewegung} on the spatial distribution of the normalized Stokes parameters characterizing the polarization of the polariton condensate.
}
\end{figure}

Although the \emph{zitterbewegung} of exciton-polaritons is clearly observed in the spatial intensity distribution, it is even better seen in the spatial distribution of the polariton polarization, characterized by the pseudo-spin (Stokes) vector $\mathbf{S}=\frac{1}{2}\langle\Psi| \oht{\boldsymbol{\sigma} }|\Psi\rangle$ with the components $S_{x}=\mathrm{Re}(\Psi_+^*\Psi_-)$, $S_{y} = \text{Im}(\Psi_+^*\Psi_-)$ and $S_{z}=(|\Psi_+|^2-|\Psi_-|^2)/2$.
Figure~\ref{FIGpolarizations} demonstrates the spatial distribution of the components of the normalized vector 
$\mathbf{s}=\left.\mathbf{S}\right/S$ in the \emph{zitterbewegung} regime in the steady state.
The effect results in the appearance of the real-space patterns resembling interlocked fingers for $s_y$ and $s_z$  components.
We note that the predicted polarization effect is expected to be easily observable experimentally due to the wider area in which it is manifested.

The \emph{zitterbewegung} can be generally characterized by the period of oscillations of the displacement $L$, the amplitude of oscillations $A$, and the decay length $\Lambda$ (the distance at which the amplitude $A$ decreases by a factor of~$e$).
To optimize the characteristics of the \emph{zitterbewegung}  trajectories we calculate them as functions of the most important parameters of the sample and of the experimental configuration.
Among them are the splitting constant $\Delta_{\mathrm{LT}}$, the wave number of the probe $k_{0}$  and the size of the condensate $d$ governed by the spatial width of the probe beam~$w_\mathrm{p}$~\cite{ValuesOfTheParameters}.

\begin{figure}
\includegraphics[width=1.\columnwidth]{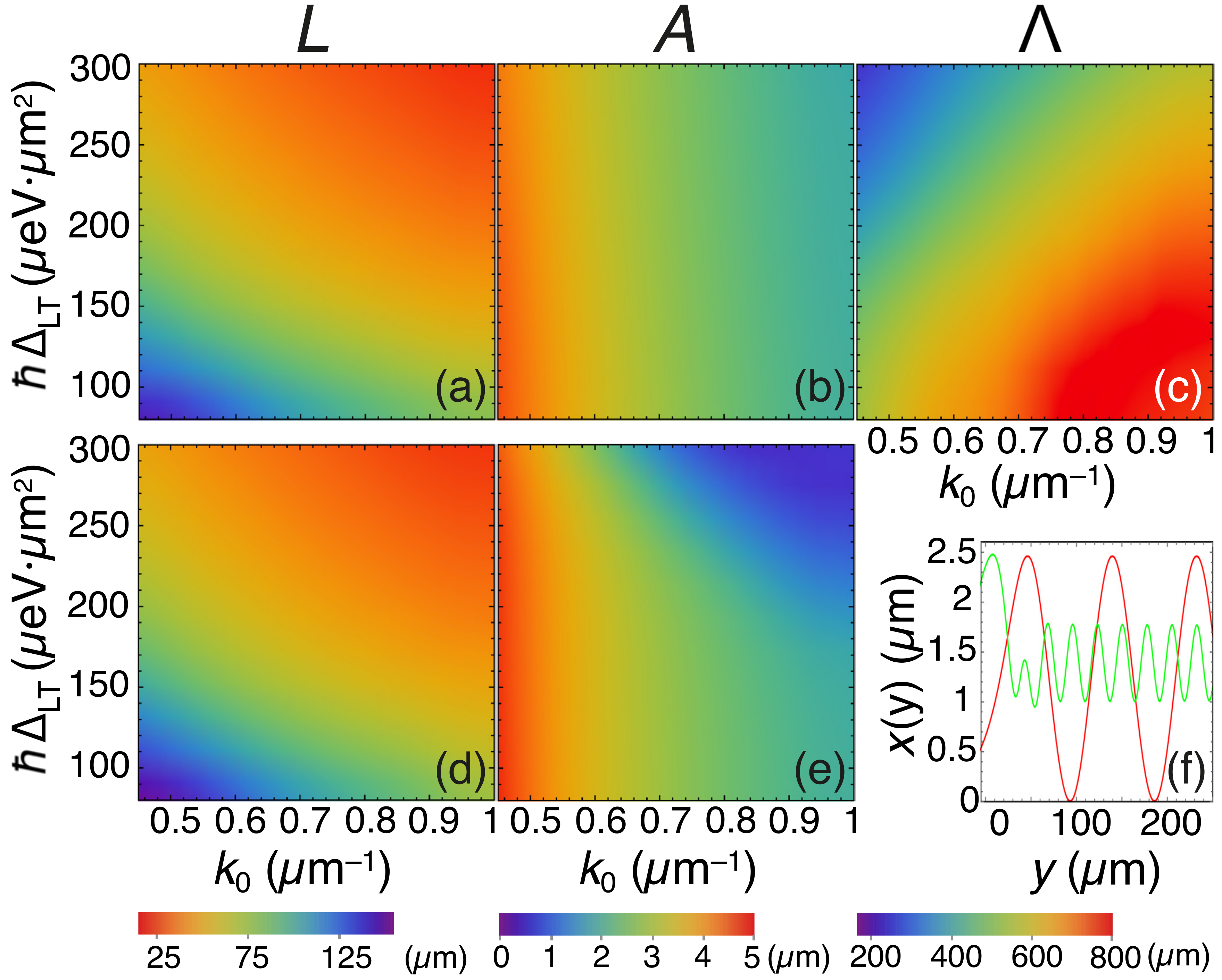}
\caption{ \label{FIGparameters}
(Color online)
The period $L$ ((a) and (d)), the amplitude $A$ ((b) and (e)) and the decay length $\Lambda$ (c) of the \textit{zitterbewegung} as functions of the LT-splitting constant $\Delta_{\text{LT}}$ and the wave number $k_{0}$ of the probe.
Upper and lower panels correspond to the different widths of the probe, $w_\mathrm{p} = 10 \, \mu \text{m}$ and $30 \, \mu \text{m}$, respectively.
The panel (f) shows variation in space of the expectation value of position of the center of mass of the polariton condensate $x(y)$ for different parameters: $\hbar \Delta _{\text{LT}} = 90 \, \mu \text{eV} \cdot \mu \text{m}^{2}$ (red) and $300 \, \mu \text{eV} \cdot \mu \text{m}^{2}$ (green).
The other parameters for (f) are: $k_{0} = 0.8 \, \mu \text{m}$ and $w_\mathrm{p} = 30 \, \mu\text{m}$.
}
\end{figure}

Color maps in Fig.~\ref{FIGparameters} demonstrate the characteristics of the \textit{zitterbewegung} as functions of the control parameters of $\Delta _{\text{LT}}$ and $k_{0}$ for two values of the probe beam width~$w_\mathrm{p}$.
Areas with the preferential from the point of view of the \textit{zitterbewegung} observation values of the parameters are shown by a red color.
Figures~\ref{FIGparameters}(a) and~\ref{FIGparameters}(d) show that the period $L$ decreases with the increase of either $\Delta_{\mathrm{LT}}$ or $k_{0}$, and both dependencies are close to hyperbolic.
The same character of the dependence is predicted by the expression~\eqref{xONyEQ}.
The probe width $w_\mathrm{p}$ hardly influences the \textit{zitterbewegung} period, as one can see from Figs.~\ref{FIGparameters}(a) and~\ref{FIGparameters}(d).

The analytical expression~\eqref{xONyEQ} describing the trajectory of polaritons, valid for large $k_0$, also predicts the hyperbolic dependence of the amplitude of the \emph{zitterbewegung} on the wave number $k_{0}$. 
In general, it is confirmed by the results of modelling for the probe beam of width not exceeding period of oscillations, see Fig.~\ref{FIGparameters}(b).
Notably, the amplitude hardly depends on the LT splitting.
For a wider probe (Fig.~\ref{FIGparameters}(e)), the amplitude still decreases with the increasing wave number but the dependence is stronger for larger LT-splitting in this case.

In all cases, the \emph{zitterbewegung} exhibits a transient character.
Figure~\ref{FIGparameters}(c) shows that the decay length $\Lambda$ decreases with the increase of $\Delta _{\text{LT}}$ and with the decrease of~$k_{0}$.
The effect of the pump width on the shape of polariton wave packet trajectory has also been investigated by us.
However we are unable to provide a diagram similar to Fig.~\ref{FIGparameters}(c) for a wider probe beam.
If the probe width $w_\mathrm{p}$ is comparable or exceeds the period $L$, it directly affects  the condensate distribution in the close to the injection spot region.
The decay length $\Lambda$ hardly can be defined in this case.
The direct effect of the probe is clearly seen in the panel Fig.~\ref{FIGparameters}(e) where the amplitude $A$ is plotted for large $\Delta _{\text{LT}}$ and $k_{0}$ corresponding to the small period compared to $w_\mathrm{p}$.
To qualitatively demonstrate the effect of the pump width, in Fig.~\ref{FIGparameters}(f) we show the spatial variation of the center of mass of the polariton condensates with different periods of the \emph{zitterbewegung} excited by the probe beam of width $w_{\text{p}}=30 \, \mu \text{m}$.
The effect of the probe on the condensate with the smaller period (green curve) results in the significantly smaller amplitude and the extremely rapid decay of the \emph{zitterbewegung} in the region close to the injection spot in comparison with the condensate with the wider period (red curve). 
Also, one should mention that once the condensate leaves the probe beam area, the \emph{zitterbewegung} amplitude becomes weakly damped with the distance.
Consequently, the effect of \emph{zitterbewegung} of exciton-polaritons can not be observed in two opposite limits of small and large LT-splittings, as either the inversed period or the amplitude vanish in these limits: $1/L \rightarrow 0$ for $\Delta _{\text{LT}} \rightarrow 0$ and $\Lambda \rightarrow 0$ for $\Delta _{\text{LT}} \rightarrow \infty$.
In the case of small $\Delta _{\text{LT}} $ the effect of \emph{zitterbewegung} results in the drift of a polariton condensate in $x$ direction until its eventual decay.
Analogously, the \emph{zitterbewegung} becomes indiscernible both for small and large wave numberes $k_{0}$. 
In the former case, although the amplitude increases, the period increases as well.
In the latter case, the amplitude drops down with the increasing $k_{0}$.
Notably, the polariton lifetime does not affect the \emph{zitterbewegung}  in the considered regime  of the sub-threshold cw pump. 

In conclusion, we have theoretically studied the \emph{zitterbewegung} of exciton-polaritons in the driven-dissipative system of a 2D microcavity characterised by the LT splitting of polariton modes.
We have shown that the condensed polaritons propagate in the cavity plane following an oscillating trajectory that appears as a result of the pseudospin rotation in the course of the motion of polariton wave packet.
The polariton polarization degree is strongly affected by the \emph{zitterbewegung}, which helps experimental observation of the effect.
We propose the optimized control parameters of the system that would help observation of the \emph{zitterbewegung} on the length scale of tens of micrometers.

\acknowledgments
E.S.S.~acknowledges support from the RFBR grants No.~16-32-60104 and 17-52-10006. 
Y.G.R.~acknowledges support from CONACYT (Mexico) grant No.\ 251808.  
A.V.K.~and E.S.S.\ acknowledge support from the Engineering and Physical Sciences Research Council (EPSRC) Programme Grant No. EP/M025330/1 ``Hybrid Polaritonics.''
A.V.K.~acknowledges the joint Russian-Greek project supported by Ministry of Education and Science of The Russian Federation (project RFMEFI61617X0085).

\bibliography{ZitterbewegungBibl}

\begin{thebibliography}{23}
\bibitem [{Val()}]{ValuesOfTheParameters}%
  \BibitemOpen
  \href@noop {} {}\bibinfo {note} {Parameters: $m^* = 7 \cdot 10^{-5} m_e$,
  where $m_e$ is the free electron mass, $\hbar \Delta _{\text{LT}} = 300
  \,\text{meV} \cdot \mu\text{m}^2$, $\gamma _C = 0.02 \, \text{ps}^{-1}$,
  $\gamma _R = 0.025 \, \text{ps}^{-1}$, $\hbar R = 0.05 \, \text{meV} \cdot
  \mu\text{m}^2$. The nonlinear coefficients are $\alpha _{1} = 5\, \mu
  \text{eV} \cdot \mu \text{m}^2$, $\alpha _2 = -0.05 \alpha _1 $ and ${\alpha
  _{\text{R}} = 2 \alpha _1}$. The non-resonant pump power is taken homogenious
  $P_{\pm} = 0.95 P_{\text{th}}$. The resonant pump is taken circularly
  polarized with the Gaussian beam cross-section: $E_{\text{p}+} \propto \exp
  \left[ -r^2 / w_\text{p} ^2 \right] $, $E_{\text{p}-} =0 $. The beam width is
  taken $w_{\text{p}}=10 \, \mu\text{m}$, the pump energy is $\hbar \omega
  _{\text{p}} = 0.196$~meV, the corresponding wave number is $k_{\text{p}} =
  0.6\, \mu\text{m}^{-1}$.}\BibitemShut {Stop}%
\end{thebibliography}

\end{document}